\documentstyle[aps,pra]{revtex} 
\begin{document}

%\draft

\twocolumn[\hsize\textwidth\columnwidth\hsize\csname
@twocolumnfalse\endcsname

\title{Band structure effects on the interaction of charged particles with solids}
\author{J. M. Pitarke$^{1,2}$ and I. Campillo$^{1}$}
\address{$^1$Materia Kondentsatuaren Fisika Saila, Zientzi Fakultatea,
Euskal Herriko Unibertsitatea,\\ 
644 Posta Kutxatila, 48080 Bilbo, Basque Country, Spain\\
$^2$Donostia International Physics Center (DIPC) and Centro Mixto CSIC-UPV/EHU,\\
Donostia, Basque Country, Spain}

\date\today

\maketitle

\begin{abstract}
A survey is presented of current investigations of the impact of band structure effects on various
aspects of the interaction of charged particles with real solids. The role that interband
transitions play in the decay mechanism of bulk plasmons is addressed, and results for
plasmon linewidths in Al and Si are discussed. {\it Ab initio} calculations of
the electronic energy loss of ions moving in Al and Si are also presented, within linear
response theory, from a realistic description of the one-electron band structure and a full
treatment of the dynamic electronic response of valence electrons. Both random and
position-dependent stopping powers of valence electrons are computed.
\end{abstract}
]

%\narrowtext

\section{Introduction}

A quantitative description of the interaction of charged particles with solids is of fundamental
importance in a variety of theoretical and applied areas\cite{Echenique}. When a moving charged
particle penetrates a solid material, it may lose energy to the medium through various elastic and
inelastic collision processes that are based on electron excitation and nuclear recoil
motion in the solid. While energy losses due to nuclear recoil may become dominant at very low
energies of the projectile\cite{Komarov}, in the case of electrons or ions moving with
non-relativistic velocities that are comparable to the mean speed of electrons in the solid the most
significant energy losses are due to the generation of electronic excitations, such as electron-hole
pairs, collective oscillations, i.e., plasmons, and inner-shell excitations and ionizations.

For many years, theoretical investigations of valence-electron excitations in metals have been
carried out on the basis of the so-called {\it jellium} model of the solid. Within this model,
valence electrons are described by a homogeneous assembly of free electrons immersed in a uniform
background of positive charge, the only parameter being the valence-electron density $n_0$ or,
equivalently, the so-called electron-density parameter $r_s$ defined by the
relation $1/n_0=(4/3)\,\pi\,(r_s\,a_0)^3$, $a_0$ being the Bohr radius. On the other hand, a widely
used approach to treat long-range Coulomb interactions in a many-electron system is the so-called
random-phase approximation (RPA)\cite{Pines}, a mean-field theory that has been proved to be
successful in describing both electron-hole and plasmon excitations. For instance, the free-electron
gas (FEG) predicts, within RPA, a quadratic plasmon dispersion $\omega(q)$ with momentum transfer
which is in qualitative agreement with many experimental results\cite{Raether}.

Among the limitations of a free-electron-gas description of the solid there is the absence, within
RPA, of any damping mechanism other than the Landau damping where plasmons can decay, for momentum
transfers above the critical value $q_c$, through the creation of an electron-hole pair. As a
result, the FEG predicts, within RPA, an infinitely sharp plasmon line for momentum
transfers smaller than $q_c$, while measured plasmon lines of real solids show a finite
linewidth\cite{Raether}.

Plasmon-decay mechanisms are due, within a free-electron-gas description of the solid, to the
existence of frequency-dependent many-body interactions beyond RPA. These high-order interactions
yield non-zero probabilities for the plasmon to decay into two electron-hole pairs or into a plasmon
of lower energy plus an electron-hole pair, thereby satisfying momentum and energy conservation for
arbitrarily small values of $\bf q$. However, this plasmon damping, which has been shown to be of
order $q^2$ for small values of the wave vector, cannot explain the observed finite width for $q\to
0$. Also, calculations by DuBois and Kivelson\cite{DuBois} and by Hasegawa and Watabe\cite{Hasegawa}
show that this contribution to the plasmon linewidth is too small to account alone for the linewidth
dispersion in simple metals.

Additional plasmon-decay mechanisms, which yield a finite plasmon linewidth at zero and non-zero
momentum transfers, are due to either phonon/impurity assisted electron-hole excitations or to
band-structure effects. The former can be accounted by simply introducing a phenomenological
relaxation time in the particle-conserving RPA-like Mermin dielectric function\cite{Mermin}. The
later requires the calculation of the wave-vector and frequency dependent dielectric matrix in the
reduced translational symmetry of the real solid. Early calculations of band-structure effects on the
plasmon lifetime in simple metals were performed by Sturm and Oliveira\cite{Sturm,Oliveira,Sturm2},
with the use of a nearly free-electron pseudopotential theory. {\it Ab initio} calculations of the
plasmon lifetime have been carried out only very recently for the case of potassium\cite{Eguiluz},
thereby explaining the experimentally observed anomalous dispersion of the plasmon linewidth in this
simple metal\cite{Fink}.

Calculations of the electronic stopping power of solids, i.e., the energy that moving charged
particles loose per unit path-length due to electronic excitations in the solid, have also been
carried out, for many years, on the basis of a {\it jellium} model of the target\cite{Ritchie}. The
calculation of the stopping power of real solids from the knowledge of the band structure and the
corresponding Bloch eigenfunctions is a laborious problem, and early theoretical investigations
were based on either semiempirical treatments of the electronic excitations in the
solid\cite{Kom,Des,Esbensen,Kom2,Crawford} or approximate schemes based on the assumption that
electrons are individually bound by harmonic forces\cite{Sigmund}. Among the most recent attempts to
introduce the full electronic band structure in the evaluation of the electronic stopping power for low
projectile velocities there is, for alkaline metals, a one-band calculation\cite{Grande}, as well as a
calculation based on a linear combination of atomic orbitals (LCAO)\cite{Dorado}. The low-velocity
limit was also investigated, in the case of silicon, on the basis of a static treatment of the
density-response of the solid\cite{Tielens}, and, more recently, {\it ab initio} band structure
calculations that are based on a full evaluation of the dynamical density-response of the solid have
been carried out for aluminum\cite{Igor0,Igor1} and silicon\cite{Igor2}.

In this paper we summarize current investigations of the impact of band structure effects on
various aspects of the interaction of charged particles with real solids. In Sec. II, we
describe our full treatment of the dynamical density-response of valence electrons which is based,
in the framework of time-dependent density-functional theory (TDDFT)\cite{tddft1,tddft2}, on a
realistic description of the one-electron band structure and first-principles pseudopotential
theory. In Sec. III, band-structure effects on the energy-loss function are investigated. For wave
vectors below the plasmon cutoff $q_c$ we focus on both the dispersion of plasmon energy and plasmon
linewidths, and for larger wave vectors we investigate the so-called dynamic structure factor. In
Sec. IV, the electronic stopping power of valence electrons in aluminum and silicon is considered,
which we evaluate within either random or channeling conditions. In Sec. V, our conclusions are
presented. In the present work, we restrict our attention to the lowest order in the projectile
charge, thereby high-order corrections\cite{z3} being ignored. Hence, existing differences between the
stopping power for protons and antiprotons are not accounted here, and we focus on the impact of band
structure effects on the stopping power of valence electrons in real solids.

\section{Linear response}

Take a system of interacting electrons exposed to an external potential
$V^{ext}({\bf r},\omega)$. According to time-dependent perturbation theory and
keeping only terms of first order in the external perturbation, the electron density induced in the
electronic system is found to be
\begin{eqnarray}\label{eqa1}
\rho^{ind}({\bf r},\omega)=\int{\rm d}{\bf r}'\,
\chi({\bf r},{\bf  r'};\omega)\,V^{ext}({\bf r'},\omega),
\end{eqnarray}
where $\chi({\bf r},{\bf  r'};\omega)$ represents the so-called linear density-response
function\cite{Pines}
\begin{eqnarray}\label{eqa2}
\chi({\bf r}, {\bf r'},\omega)=\sum_n \, && 
\rho^{*}_{n0}({\bf r}) \rho_{n0}({\bf r}') 
\left[ {1\over E_0-E_n+\hbar(\omega+{\rm i}\eta)} \nonumber \right. \\
&& \left. -{1\over E_0+E_n+\hbar(\omega+{\rm i}\eta)}\right],
\end{eqnarray}
$\eta$ being a positive infinitesimal. $\rho_{n0}({\bf r})$ represent matrix elements, taken
between the unperturbed many-particle ground state $|\Psi_{0}\rangle$ of energy $E_0$ and the
unperturbed many-particle excited state $|\Psi_{n}\rangle$ of energy $E_n$, of the particle-density
operator
\begin{equation}\label{eqa4}
\rho({\bf r})=\sum_{i=1}^N\delta({\bf r}-{\bf r}_i),
\end{equation}
where ${\bf r}_i$ describe electron coordinates. The inverse dielectric function is connected with
the density-response function by the following relation:
\begin{equation}\label{epsilon}
\epsilon^{-1}({\bf r},{\bf r}';\omega)=\delta({\bf r}-{\bf r}')+\int{\rm d}{\bf r}''v({\bf r}-{\bf
r}'')\chi({\bf r}'',{\bf r}';\omega),
\end{equation}
where $v({\bf r}-{\bf r}')$ is the bare Coulomb interaction. 

In the framework of TDDFT, the exact density-response function $\chi({\bf r},{\bf 
r'};\omega)$ obeys the integral equation\cite{tddft1}
\begin{eqnarray}\label{eq98p}
\chi({\bf r},&&{\bf r}';\omega)=\chi^0({\bf r},{\bf r}';\omega)
+\int{\rm d}{\bf r}_1\int{\rm d}{\bf r}_2\,\chi^0({\bf r},{\bf r}_1;\omega)\cr\cr
&&\times\left[v({\bf r}_1-{\bf r}_2)
+K^{xc}({\bf r}_1,{\bf r}_2;\omega)\right]
\chi({\bf r}_2,{\bf r}';\omega),
\end{eqnarray}
where $\chi^0({\bf r},{\bf r}';\omega)$ represents the density-response function for
{\it non-interacting} Kohn-Sham electrons\cite{dft},
and $K^{xc}({\bf r},{\bf r}';\omega)$ represents the reduction in the electron-electron interaction
due to the existence of dynamical short-range exchange and correlation (XC) effects.

In the case of a homogeneous electron gas, one introduces Fourier transforms and writes
\begin{equation}\label{eq28}
\chi_{{\bf q},\omega}=\chi_{{\bf q},\omega}^0+\chi_{{\bf 
q},\omega}^0\left(v_{\bf
q}+K^{xc}_{{\bf q},\omega}\right)\chi_{{\bf q},\omega},
\end{equation}
where $v_{\bf q}$ represents the Fourier transform of the bare Coulomb interaction, $\chi_{{\bf
q},\omega}^0$ is the well-known function of Lindhard\cite{Lindhard}, and
\begin{equation}\label{eq29}
K^{xc}_{{\bf q},\omega}=-v_{\bf q}\,G_{{\bf q},\omega},
\end{equation}
$G_{{\bf q},\omega}$ being the so-called
local-field factor\cite{Singwi,Ichimaru,Gold}.
 
For a periodic crystal, we
introduce the following Fourier expansion of the density-response function:
\begin{equation}\label{eq8}
\chi({\bf r},{\bf r}';\omega)={1\over V}\sum_{\bf q}^{BZ}\sum_{{\bf
G},{\bf G}'}{\rm e}^{{\rm i}({\bf q}+{\bf G})\cdot{\bf r}}{\rm
e}^{-{\rm i}({\bf q}+{\bf G}')\cdot{\bf r}'}\chi_{{\bf G},{\bf G}'}({\bf
q},\omega),
\end{equation}
where $V$ represents the normalization volume, the first sum runs over ${\bf q}$ vectors within
the first Brillouin zone (BZ), and ${\bf G}$ and ${\bf G}'$ are reciprocal lattice
vectors. Introduction of Eq. (\ref{eq8}) into Eqs. (\ref{epsilon}) and (\ref{eq98p}) yields matrix
equations for the Fourier coefficients $\chi_{{\bf G},{\bf G}'}({\bf q},\omega)$ and
$\epsilon^{-1}_{{\bf G},{\bf G}'}({\bf q},\omega)$. In the case of {\it non-interacting}
Kohn-Sham electrons, one finds:
\begin{eqnarray}\label{eq9}
&&\chi_{{\bf G},{\bf G}'}^0({\bf q},\omega)={1\over V}\sum_{\bf
k}^{BZ}\sum_{n,n'} {f_{{\bf k},n}-f_{{\bf k}+{\bf q},n'}\over E_{{\bf
k},n}-E_{{\bf k}+{\bf q},n'} +\hbar(\omega + {\rm i}\eta)}\cr\cr
&&\times\langle\phi_{{\bf k},n}|e^{-{\rm i}({\bf q}+{\bf G})\cdot{\bf
r}}|\phi_{{\bf k}+{\bf q},n'}\rangle
\langle\phi_{{\bf k}+{\bf q},n'}|e^{{\rm i}({\bf q}+{\bf G}')\cdot{\bf
r}}|\phi_{{\bf k},n}\rangle.
\end{eqnarray}
The second sum runs over the band structure for each wave vector ${\bf k}$ in the first
Brillouin zone, $f_{{\bf k},n}$ represents the Fermi-Dirac distribution function, and $\phi_{{\bf
k},n}({\bf r})$ and
$E_{{\bf k},n}$ are Bloch eigenfunctions and eigenvalues of the Kohn-Sham Hamiltonian of
density-functional theory (DFT)\cite{dft}, which we evaluate within the local-density
approximation (LDA) with use of the parametrization of Perdew and Zunger\cite{Perdew}. All results
presented in this article have been found to be well converged with the
use in Eq. (\ref{eq9}) of 60 (Al) and 200 (Si) bands. The sum over the
Brillouin zone has been performed with the use of $10\times 10\times 10$ and $8\times 8\times 8$
Monkhorst-Pack meshes\cite{Pack} for Al and Si, respectively. 

For the evaluation of the one-electron $\phi_{{\bf k},n}({\bf r})$ eigenfunctions, we first
expand them in a plane-wave basis,
\begin{equation}\label{eq10}
\phi_{{\bf k},n}({\bf r})={1\over\sqrt V}\sum_{\bf G} u_{{\bf k},n}({\bf
G}){\rm e}^{{\rm i}({\bf k}+{\bf G})\cdot{\bf r}},
\end{equation}
with a kinetic-energy cutoff that varies from 12 Ry in the case of Al ($\sim 100\,{\bf
G}$-vectors) to 16 Ry in the case of Si ($\sim 300\,{\bf G}$-vectors). Then,
we evaluate the coefficients $u_{{\bf k},n}$ by solving the Kohn-Sham equation of DFT with a full
description of the electron-ion interaction that is based on the use of an {\it ab initio} non-local,
norm-conserving ionic pseudopotential\cite{Hamann}. Finally, we evaluate from Eq. (\ref{eq9}) the
Fourier coefficients of the {\it non-interacting} density-response function, and solve a matrix
equation for the Fourier coefficients of the {\it interacting} density-response function, which we
obtain within either RPA or TDLDA. 

In the RPA, the XC kernel $K^{xc}({\bf r},{\bf r}';\omega)$ entering Eq. (\ref{eq98p}) is set equal
to zero. Within TDLDA, which represents an adiabatic extension to finite frequencies of the
LDA, one writes
\begin{equation}\label{eq7}
K_{xc}^{LDA}({\bf r},{\bf r}';\omega)=\delta({\bf r}-{\bf r}')\left [{{\rm d}
V_{xc}(n)\over{\rm d}n}\right]_{n=n_0({\bf r})}, 
\end{equation}
where $V_{xc}(n)$ represents the LDA XC potential entering the Kohn-Sham equation of DFT, and
$n_0({\bf r})$ is the actual density of the electron system. In the case of a homogeneous electron
gas of density
$n_0$, the Fourier transform of the XC kernel of Eq. (\ref{eq7}) can be written in the form of Eq.
(\ref{eq29}) with
\begin{equation}\label{eq32}
G_{{\bf q},\omega}^{LDA}=A\left({q\over q_F}\right)^2,
\end{equation}
where
\begin{equation}\label{eq33}
A={1\over 4}-{4\pi a_0^2\over e^2q_F^2}\,\left[{{\rm d}V_{c}(n)\over{\rm d}n}\right]_{n=n_0},
\end{equation}
$V_c(n)$ being the correlation contribution to $V_{xc}$, and $q_F$, the Fermi momentum.

\section{Energy-loss function}

\subsection{Theory}

The Hamiltonian describing the interaction between a particle of charge $Z_1e$ at point ${\bf r}$ and
the many-electron system is given by
\begin{equation}\label{eq48}
H_I=-Z_1e^2\int d{\bf r}'{\rho({\bf
r}')\over|{\bf r}-{\bf r}'|},
\end{equation}
where $\rho({\bf r})$ represents the particle-density operator of Eq. (\ref{eqa4}).

Within
first-order perturbation theory, the probability per unit time for the probe particle to scatter
from an initial state $|i>$ of energy $\varepsilon_i$ to a final state $|f>$ of energy
$\varepsilon_f$, by carrying the Fermi gas from the many-particle ground state
$|\Psi_0>$ of energy $E_0$ to some excited many-particle state $|\Psi_n>$ of energy $E_n$, is given
by the following expression:
\begin{equation}\label{eq49}
P_{i\to
f}={2\pi\over\hbar}\,\sum_n\left|<\Psi_n\,f|H_I|\Psi_0\,i>\right|^2
\delta(\varepsilon_i-\varepsilon_f+E_0-E_n).
\end{equation}

If one chooses the probe particle to be described by plane-wave states
\begin{equation}\label{eq19}
\phi_i({\bf r})={1\over\sqrt V}{\rm e}^{{\rm i}{\bf q}_i\cdot{\bf r}}
\end{equation}
and
\begin{equation}\label{eq20}
\phi_f({\bf r})={1\over\sqrt V}{\rm e}^{{\rm i}{\bf q}_f\cdot{\bf
r}},
\end{equation}
one finds the probability per unit time for the probe
particle to transfer momentum $\hbar{\bf q}=\hbar({\bf q}_i-{\bf q}_f)$ to the Fermi gas to be given
by the following expression:
\begin{equation}\label{eq51} 
P_{\bf q}={2\,\pi\over(\hbar V)^2}\,Z_1^2\,v_{\bf q}^2\,S({\bf q},\omega),
\end{equation}
where $\hbar\omega$ represents the energy transfer,
\begin{equation}
\hbar\omega={\hbar^2\over 2m_e}\left(k_i^2-k_f^2\right),
\end{equation}
and $S({\bf q},\omega)$ is the so-called dynamic structure factor of the many-electron
system:
\begin{equation}
S({\bf q},\omega)=\sum_n\left|(\rho_{\bf q}^\dagger)_{n0}\right|^2
\delta(\omega-\omega_{n0}).
\end{equation}
Here, $\rho_{\bf q}$ represents the Fourier transform of the particle density, $(\rho_{\bf
q}^\dagger)_{n0}$ are matrix elements taken between the many-particle eigenstates $|\Psi_0>$ and
$|\Psi_n>$, and $\hbar\omega_{n0}=E_n-E_0$.

The double differential cross section ${\rm d}^2\sigma/{\rm d}\Omega{\rm d}\omega$ for inelastic
scattering of charged particles is simply the number of particles being scattered per unit time,
unit solid angle, and unit frequency into the solid angle
$\Omega$ with energy transfer $\hbar\omega$, divided by the initial particle flux. From Eq.
(\ref{eq51}), one easily finds
\begin{equation}
{{\rm d}^2\sigma\over{\rm d}\Omega{\rm d}\omega}={Z_1^2\over 4\pi^2a_0^2e^4}{q_f\over q_i}v_{\bf
q}^2\,S({\bf q},\omega).
\end{equation}  

While high-energy transmission electron beams have been used to probe $S({\bf q},\omega)$ for
$q<q_F$, thereby providing experimental evidence of collective excitations\cite{Raether}, higher
values of
${\bf q}$ have been studied with use of inelastic X-ray scattering\cite{Platzman}. Within the
lowest-order Born approximation, the inelastic X-ray scattering cross section is given by
\begin{equation}
{{\rm d}^2\sigma\over{\rm d}\Omega{\rm d}\omega}=({\bf e}_0\cdot{\bf
e}_1)^2\pi r_0^2\,{\omega_1\over\omega_0}\,S({\bf q},\omega),
\end{equation}
where $({\bf e}_0,\omega_0)$ and $({\bf e}_0,\omega_0)$ are the polarization and frequency of the
incident and scattered photon, respectively, and $r_0$ is the classical electron radius.

The dynamic structure factor, which determines the fluctuation of the particle density, is
connected to the imaginary part of the density-response function of Eq. (\ref{eqa2}) through the
fluctuation-dissipation theorem. At zero temperature, one writes\cite{Doniach}:
\begin{equation}\label{fd}
S({\bf q},\omega)=-{\hbar\over\pi}\int{\rm d}{\bf r}\int{\rm d}{\bf r}'{\rm e}^{-{\rm i}\,{\bf
q}\cdot({\bf r}-{\bf r}')}{\rm Im}\,\chi({\bf r},{\bf r}';\omega).
\end{equation}
Hence, in the case of a uniform electron gas one easily finds the absorption probability
$P_{\bf q}$ to be given by the following expression:
\begin{equation}\label{h}
P_{{\bf q}}={2\over\hbar V}\,Z_1^2\,v_{\bf q}\,{\rm
Im}\left[-\epsilon^{-1}({\bf q},\omega)\right],
\end{equation}
where ${\rm Im}\left[-\epsilon^{-1}({\bf q},\omega)\right]$ is the so-called energy-loss function
of a homogeneous electron gas. For periodic crystals, introduction of Eq. (\ref{eq8}) into Eq.
(\ref{fd}) yields:
\begin{equation}\label{old}
P_{{\bf q}+{\bf G}}={2\over\hbar V}\,Z_1^2\,v_{{\bf q}+{\bf G}}\,{\rm
Im}\left[-\epsilon_{{\bf G},{\bf G}}^{-1}({\bf q},\omega)\right],
\end{equation}
where ${\bf q}$ represents a wave vector in the first BZ, $\hbar({\bf q}+{\bf G})$ is the
momentum transfer, and $\epsilon_{{\bf G},{\bf G}'}^{-1}({\bf q},\omega)$ is
the inverse dielectric matrix:
\begin{equation}\label{eq27}
\epsilon_{{\bf G},{\bf
G}'}^{-1}({\bf q},\omega)=\delta_{{\bf G},{\bf G}'}+v_{{\bf q}+{\bf G}}\,\chi_{{\bf G},{\bf G}'}
({\bf q},\omega).
\end{equation}
If recoil of the probe particle can be neglected, as in the case of heavy projectiles or fast
electrons with small values of the momentum transfer, the energy transfer is simply
$\hbar\omega=\hbar{\bf q}\cdot{\bf v}$. 

We note that as long as the probe particle can be described by plane waves, the absorption
probability in periodic crystals is proportional to the imaginary part of one diagonal element of
the energy-loss matrix ${\rm Im}\left[-\epsilon^{-1}_{{\bf G},{\bf G}'}({\bf
q},\omega)\right]$. Hence, crystalline local-field effects\cite{Adler,Wiser}, i.e., couplings between
${\bf q}+{\bf G}$ and ${\bf q}+{\bf G}'$ wave vectors with ${\bf G}\neq{\bf G}'$ only enter through
the dependence of the diagonal elements of the inverse dielectric matrix on the off-diagonal
elements of the direct matrix.

When the probe particle moves with constant velocity ${\bf v}$ on a
definite trajectory at a given impact vector ${\bf b}$, initial and final states can be described in
terms of plane waves in the direction of motion and a
$\delta$ function in the transverse direction\cite{Igor1}. As a result, the probability per
unit time for the probe particle to transfer momentum $\hbar{\bf q}$ to the Fermi gas is now given by
the following expression:
\begin{equation}\label{eq51p} 
P_{\bf q}={2\,\pi\over(\hbar V)^2}\,Z_1^2\,\sum_{{\bf q}'}\,v_{\bf q}\,v_{{\bf
q}'}\,{\rm e}^{{\rm i}{\bf b}\cdot({\bf q}+{\bf q}')}\,S({\bf q},{\bf
q}';\omega)\,\delta_{q_z-q_z'},
\end{equation}
where $\omega={\bf q}\cdot{\bf v}$, $\delta_{q_z-q_z'}$ represents the Kroenecker $\delta$
symbol, $q_z$ and $q_z'$ are components of ${\bf q}$ and ${\bf q}'$ in the direction of motion, and
\begin{equation}\label{fdp}
S({\bf q},{\bf q}';\omega)=-{\hbar\over\pi}\int{\rm d}{\bf r}\int{\rm d}{\bf r}'{\rm e}^{-{\rm
i}\,({\bf q}\cdot{\bf r}+{\bf q}'\cdot{\bf r}')}{\rm Im}\,\chi({\bf r},{\bf
r}';\omega).
\end{equation}

In the case of a homogeneous electron gas the probability $P_{\bf q}$ of Eq. (\ref{eq51p}) is easily
found to be independent of the impact vector and to coincide with that of Eq. (\ref{h}).
However, for periodic crystals introduction of Eq. (\ref{eq8}) into Eq. (\ref{fdp}) yields the
position-dependent absorption probability:
\begin{equation}\label{new}
P_{{\bf q}+{\bf G}}={2\over\hbar V}Z_1^2\sum_{\bf K}{'}\,{\rm e}^{{\rm
i}{\bf K}\cdot{\bf b}}\,v_{{\bf q}+{\bf G}+{\bf K}}\,{\rm
Im}\left[-\epsilon_{{\bf G},{\bf G}+{\bf K}}^{-1}({\bf q},\omega)\right],
\end{equation}
where the prime in the summation indicates that it is restricted to those reciprocal-lattice
vectors that are perpendicular to the velocity of the projectile, i.e., ${\bf K}\cdot{\bf v}=0$.
The most important contribution to the position-dependent probability of Eq. (\ref{new}) is
provided by the term ${\bf K}=0$, the magnitude of higher-order terms depending on the direction of
the velocity. For those directions for which the condition ${\bf K}\cdot{\bf v}=0$ is never
satisfied we have the absorption probability of Eq. (\ref{old}), and for a few highly symmetric or
channeling directions non-negligible corrections to the {\it random} result are expected. We also
note that the average over impact parameters of the position-dependent probability of Eq.
(\ref{new}) along any given channel coincides with the absorption probability of Eq. (\ref{old}).

\subsection{Results and discussion}

First of all, we examine the dispersion of the plasmon energy and linewidth by calculating the
electron energy-loss function ${\rm Im}\left[-\epsilon^{-1}_{{\bf
G},{\bf G}}({\bf q},\omega)\right]$ for wave vectors that are below the plasmon cutoff $q_c$. For
these wave vectors the FEG predicts, within either RPA or TDLDA, sharp plasmon lines, which would
be broadened through the introduction of a finite value of $\eta$ in Eq. (\ref{eq9}). In order to
avoid numerical broadening, we have computed the energy-loss
function of the real solid for imaginary frequencies and have obtained it on the real axis with
$\eta\to 0^+$ by using Pad\'e\, approximants\cite{Pade,Lee0}. 

Our computed RPA plasmon peaks in Al are displayed in Fig. 1, for increasing magnitude of the wave
vector in the (100) direction. Vertical lines are $\delta$ functions corresponding to plasmon
excitation in a FEG. As a result of the actual band structure of the solid, the energy-loss
function exhibits a sharp but finite plasmon peak at low wave vectors with a full width at half
maximum (FWHM) that increases rapidly, for increasing ${\bf q}$, while the peak height decreases
according to the $f$-sum rule. For periodic crystals, one writes:
\begin{equation}\label{sum}
\int_0^{\infty}{\rm d}\,\omega\,\omega{\rm Im}\left[-\epsilon_{{\bf G},{\bf G}'}^{-1}({\bf
q},\omega)\right]={2\pi^2e^2\over m_e}\,n_{{\bf G}-{\bf G}'},
\end{equation}
where $n_{{\bf G}}$ represents the Fourier components of the density, and $n_{{\bf G}=0}$ equals
the average electron density of the crystal.

RPA and TDLDA plasmon peak positions in Al, as calculated along the (100) direction, are plotted in
Fig. 2 by solid and dotted lines, respectively, as a function of the magnitude of the wave vector.
The calculated plasmon dispersions agree with those of previous calculations\cite{Quong,Lee}. For
comparison, RPA and TDLDA plasmon dispersions corresponding to a FEG with $r_s=2.07$ are represented
by dashed and dashed-dotted lines, respectively, showing that band-structure corrections result in a
nearly-q-independent downward shift of plasmon energy. If short-range XC effects, which reduce the
plasma frequency for large values of the wave vector, are included within the TDLDA, then one finds a
result for the plasmon dispersion that is in excellent agreement
with experiment\cite{Fink2}. 

The plasmon lifetime is defined as the inverse of the FWHM of the energy-loss peak,
$\Delta E_{1/2}$. RPA and TDLDA calculations of $\Delta E_{1/2}$ along the (100) direction in Al are
exhibited in Fig. 3, as a function of the magnitude of the wave vector. In the limit as
$q\to 0$ we find $\Delta E_{1/2}(0)\sim0.2\,{\rm eV}$, below the early calculation carried out by
Sturm with use of the nearly free electron approximation\cite{Sturm}. As for plasmon decay from the
creation of two electron-hole pairs, which is not included in our band-structure calculation, one
finds to lowest-order in a small-$q$ expansion\cite{Sturm2}:
\begin{equation}\label{mb}
\Delta E_{1/2}^{pair-pair}(q)=b\,(q/q_F)^2\,\hbar\omega_p,
\end{equation}
where $b\sim 0.03$ for $r_s\sim 2$. Adding band-structure effects, as derived from our calculated
RPA plasmon peaks, and many-body corrections, as obtained from Eq. (\ref{mb}), we have
found the results plotted in Fig. 3 by open triangles. Solid circles
correspond to the result of not including plasmon decay from many-body effects, thereby showing that
these effects give only a minor contribution to the plasmon linewidth. As phonons are
expected to give a relative contribution to
$\Delta E_{1/2}(q)$ that changes little with the magnitude of the wave vector\cite{Sturm2},
the reduced plasmon linewidth
$\Delta E_{1/2}(q)/\Delta E_{1/2}(0)$ is plotted in Fig. 4 by open circles (RPA) and triangles
(TDLDA), showing a reasonable agreement with the experimental results of Gibbons {\it
et al}\cite{Gibbons}.

In Fig. 5 we show full-band-structure calculations of RPA and TDLDA energy-loss functions in Si,
as obtained along the (111) direction. As in Fig. 1, vertical lines represent $\delta$ functions
corresponding to plasmon excitations in a FEG, now with $r_s=2.01$. Besides the finite width of
the energy-loss function of real Si, which is found to be about 5 times wider than in Al, our
calculated plasmon energy shows a characteristic $q$ dependence which is considerably different
from that of a FEG.

The energy positions of RPA and TDLDA plasmon peaks of Si are plotted in Fig. 6 along the (111)
direction, as a function of the magnitude of the wave vector, together with the FEG
prediction and the experimental result\cite{Stiebling}. In the long-wavelength limit ($q\to 0$), the
plasmon energy is expected to be only slightly larger than in the case of a FEG, in agreement with
the Penn model dielectric function\cite{Penn} for semiconductors which predicts a plasmon
frequency $\omega=(\omega_g^2+\omega_p^2)^{1/2}$, $\hbar\omega_g$ being the energy of the effective
band gap and $\omega_p$ representing the plasma frequency of the corresponding FEG. We also note
that while in the case of Al band-structure effects result in a nearly-q-independent downward
shift of the FEG curve, interband transitions in Si flatten the plasmon curve with a
considerable reduction of the plasma frequency at large values of the wave vector. Though our
calculated plasmon dispersion in Si is at low wave vectors in good agreement with the experiment, at
large $|{\bf q}|$ our predictions start to deviate from the experimental value, a result
also found in previous calculations\cite{Farid,Forsyth}.

In Fig. 7 we show our calculation of the plasmon linewidth dispersion of Si, as obtained along the
(111) direction with no inclusion of crystalline local-field effects. In the limit as $q\to 0$ we
find $\Delta E_{1/2}(0)\sim 1\,{\rm eV}$, well below the prediction of Louie {\it et
al}\,\cite{Louie} who performed a calculation of the energy-loss function in Si at ${\bf q}=0$ with
full inclusion of local-field effects. This shows the key role that coupling between ${\bf q}+{\bf
G}$ and
${\bf q}+{\bf G}'$ wave vectors with ${\bf G}\neq{\bf G}'$ plays in the plasmon-decay mechanism in
Si\cite{Igor2}, which is a consequence of the presence of directional covalent bondings in this
material. Considerable broadening of plasmon peaks in Si by the effect of local-field effects has
been reported recently by Lee and Chang\cite{Lee0}.

Next, we calculate the energy-loss function in Al and Si for large wave vectors of
$|{\bf q}+{\bf G}|=1.65q_F$ and $|{\bf q}+{\bf G}|=1.95q_F$, respectively, along the directions
(1,5/6,5/6) and (1,1,1). The results we have obtained within RPA and TDLDA are plotted in Figs. 8
and 9, showing in the case of Al the experimentally determined double-peak structure\cite{Platzman}
which was recently explained\cite{Fleszar} as originated from band-structure effects.

We note that splitting of the band structure of Al over the Fermi level opens new channels for the
creation of electron-hole pairs, which results in an energy-loss function that is, at low
frequencies, higher than in the case of a FEG (see Fig. 8). In the case of Si, the presence of the
band gap reduces the energy-loss function (see Fig. 9). As a consequence, both the absorption
probability and the electronic stopping power of Si will be, at low energies, smaller than those of
Al, though they both have about the same valence electron density.

\section{Electronic stopping power}

The electronic stopping power of a moving particle of charge $Z_1e$ is simply the energy that it
looses per unit path-length due to electronic excitations in the solid. Hence, one writes
\begin{equation}\label{eq15}  
-{{\rm d}E\over{\rm d}x}={1\over v}\sum_{\bf q}\left[\hbar\omega\,P_{\bf q}\right], 
\end{equation}
where $P_{\bf q}$ represents the probability per unit time for the projectile to transfer momentum
$\hbar{\bf q}$ to the electron gas, the sum is extended over all momentum transfers, ${\bf v}$ is
the projectile velocity, and $\hbar\omega$ is the energy transfer. As long as recoil can be
neglected, $\hbar\omega=\hbar{\bf q}\cdot{\bf v}$.

In the case of a homogeneous FEG the probability $P_{\bf q}$ is given by Eq. (\ref{h}), and
introduction of this expression into Eq. (\ref{eq15}) yields the well-known formula for the stopping
power of a FEG\cite{Echenique}. For periodic crystals, the energy loss depends on whether the
projectile is moving at a definite impact parameter or not. For random trajectories, introduction of
Eq. (\ref{old}) into Eq. (\ref{eq15}) results in the so-called random stopping power:
\begin{eqnarray}\label{eq26r}
\left[-{{\rm d}E\over {\rm d}x}\right]_{\rm random}=&&{Z_1^2\over 4\pi^3v}\,\int_{\rm
BZ}{\rm d}{\bf q}\sum_{\bf G}\,\omega\,v_{{\bf q}+{\bf G}}\cr\cr
&&\times{\rm Im}
\left[-\epsilon_{{\bf G},{\bf G}}^{-1}({\bf q},\omega)\right],
\end{eqnarray}
where the integral runs over ${\bf q}$ vectors within the first Brillouin zone, and $\omega=({\bf
q}+{\bf G})\cdot{\bf v}$.

In the case of charged particles moving with constant velocity ${\bf v}$ on a definite trajectory
at a given impact parameter ${\bf b}$, the probability per unit time for the projectile to transfer
momentum $\hbar({\bf q}+{\bf G})$ to the electron gas is given by Eq. (\ref{new}), and introduction
of this expression into Eq. (\ref{eq15}) yields the following result for the position-dependent
stopping power:
\begin{eqnarray}\label{eq26r2}
\left[-{{\rm d}E\over {\rm d}x}\right]_{\bf b}=&&{Z_1^2\over 4\pi^3v}\,\int_{\rm
BZ}{\rm d}{\bf q}\sum_{\bf G}\sum_{\bf K}{'}\,\omega\,{\rm e}^{{\rm
i}{\bf K}\cdot{\bf b}}\cr\cr
&&\times v_{{\bf q}+{\bf G}+{\bf K}}\,{\rm Im}
\left[-\epsilon_{{\bf G},{\bf G}+{\bf K}}^{-1}({\bf q},\omega)\right],
\end{eqnarray}
where the sum $\sum_{\bf K}^{'}$ is restricted, as in Eq. (\ref{new}), to those reciprocal-lattice
vectors that are perpendicular to the projectile velocity, and $\omega=({\bf q}+{\bf G})\cdot{\bf
v}$, as in Eq. (\ref{eq26r}).  

The main ingredient in the calculation of both random and position-dependent stopping powers is
the inverse dielectric matrix $\epsilon_{{\bf G},{\bf G}'}^{-1}({\bf q},\omega)$, which has
been discussed in the previous section. As the maximum energy that the moving particle
may transfer to the target is $(\hbar\omega)_{max}=q\,v$, the number of bands that are required in
the evaluation of the polarizability  $\chi_{{\bf G},{\bf G}'}^0({\bf q},\omega)$ of Eq. (\ref{eq9})
depends on the projectile velocity. Well-converged results have been found for all projectile
velocities under study, with the use in Eq. (\ref{eq9}) of 60 bands for Al and 200 bands for Si.
The sums over reciprocal-lattice vectors in Eqs. (\ref{eq26r}) and (\ref{eq26r2}) have been
extended over $15$ values of reciprocal-lattice $\bf G$ vectors, the magnitude of the maximum momentum
transfer being
$2.9q_F$ and $2.1q_F$ for Al and Si, respectively. All calculations have been performed with full
inclusion of crystalline local-field effects, i.e., by inversion of the full dielectric matrix
$\epsilon_{{\bf G},{\bf G}'}({\bf q},\omega)$, and contributions from these
so-called local-field effects to the random stopping power of Al and Si have been found to be within
$0.5\%$ and $1\%$, respectively.

\subsection{Random stopping power} 

In Fig. 10 we show, as a function of the projectile velocity, our full RPA and TDLDA results for
the random stopping power of valence electrons in real Al for protons ($Z_1=1$), together with
the corresponding result for the stopping power of a FEG with an electron-density parameter equal
to that of Al ($r_s=2.07$)\cite{note}. These results have been found to be insensitive to the choice of
the projectile-velocity direction.

As the energy-loss function of real Al is, at low frequencies, slightly enhanced with respect to the
corresponding FEG calculation (see Fig. 8), the stopping power of the real target is, for
projectile velocities smaller than the Fermi velocity and within both RPA and TDLDA, higher than
that of a FEG by about $7\%$.

At velocities over the plasmon-threshold velocity for which plasmon excitation
becomes possible, contributions to the stopping power come from both plasmon and electron-hole
excitations. These contributions have been calculated separately\cite{Igor1}, showing that
contributions from losses to plasmon excitation are independent of the detailed band structure
of the crystal. As for the contribution from the excitation of electron-hole
pairs, band-structure effects in Al are found to lower the stopping power of electrons in a FEG by
about $10\%$ at and just above the plasmon-threshold velocity.

At high velocities, well above the stopping maximum, the sum
over the frequency $\omega$ in Eq. (\ref{eq26r}) can be replaced by an integration from $0$ to
$\infty$, and the sum rule of Eq. (\ref{sum}) results in a stopping power which depends on
the average electron density $n_0$ and not on the details of the band structure of the target
material:
\begin{equation}
\left[-{{\rm d}E\over {\rm d}x}\right]_{\rm random}\sim{4\pi Z_1^2 e^4\over m_e
v^2}\,n_0\,\ln{2m_ev^2\over\hbar\omega_p}.
\end{equation}

While at low velocities the contribution to the total energy loss due to excitation of inner-shell
electrons is negligible small, at velocities larger than the Fermi velocity it is necessary to allow
for this contribution. The cross sections for the ionization of inner shells
in Al were obtained by Ashley {\it et al\,}\cite{Ashley} in the first Born approximation utilizing
atomic generalized oscillator strength functions. By adding the contribution from core electrons to
that of valence electrons (this contribution does not depend, at these velocities, on the details of
the band structure) these authors found a nice agreement with experiment. Good agreement with
experiment was also shown in Ref.\onlinecite{Komarov} by adding the energy loss from core electrons in
Si, as taken from Walske's calculations\cite{Walske}, to that from valence electrons. 

Our full calculation of RPA and TDLDA stopping powers of valence electrons in Si is plotted in Fig.
11, as a function of the projectile velocity, with $Z_1=1$, together with the corresponding result
for the stopping power of a FEG with $r_s=2.01$\cite{note}. As in the case of Al, these results have
been found to be insensitive to the choice of the direction of the projectile velocity.

The band gap in Si clearly makes the stopping power of this material different from that of a
simple metal as Al. Though the average band gap of Si is small compared to the bandwidth of both
valence and conduction bands, there are fewer low-energy excitation levels available than in the
case of a FEG, and a lower energy loss is expected.  Our calculations show that
at low velocities the stopping power of Si is smaller than that of a FEG with $r_s=2.01$ by about
$10\%$. Also, the stopping power of Si is found not to be,
at very small projectile velocities, proportional to the velocity,
in agreement with experimental low-velocity stopping
powers of this material\cite{Ziegler}.

As in the case of Al, contributions from plasmon excitation to the stopping power of Si are found
to be insensitive to the band structure, while contributions from electron-hole excitation yield a
stopping power of this material which is just over the plasmon threshold about $20\%$ lower than that
of a FEG. At high velocities, well above the stopping maximum, all calculations are found
to converge.

\subsection{Position-dependent stopping power}

We have carried out, from Eq. (\ref{eq26r2}), calculations of the position-dependent electronic
energy loss of protons in Al and Si. In the case of Al, calculations have been performed for
slow ions moving along the (100) and (111) directions\cite{Igor1}, showing that the existence of
small electron-density variations in this material result, through the presence of non-negligible
off-diagonal elements in the interacting density-response matrix, in position-dependent
corrections to the random stopping power of up to $10\%$ and $20\%$, respectively. The maxima in
the stopping power for trajectories along the interstitial regions and the minima near the cores
are associated with corresponding maxima and minima in the integrated
electronic densities along the projectile trajectories. Nevertheless, a local-density approximation
(LDA), according to which the position-dependent stopping power is obtained as that of a homogeneous
electron gas with an electron density equal to the average electron density along the projectile
trajectory, is found to predict corrections to the random stopping power which are, for
low projectile velocities, too small.

A contour density plot of the square lattice containing both the projectile trajectory along the
(110) direction in Si and the impact vector ${\bf b}$ is displayed in Fig. 12. The integrated density
in the (110) channel varies from $r_s=1.49$ at the atomic row (full diagonal of the figure) to
$r_s=3.37$ at the center of the channel (solid line with an arrow), thereby showing valence-electron
density variations of up to $80\%$ typical of covalent crystals like silicon.

In Fig. 13 we show, as a function of the projectile velocity, our full RPA calculation of the
stopping power of Si for best channeled ions moving along the (110) direction. The stopping power of
valence electrons in Si for these channeled ions is found to be, at intermediate velocities,
about $20\%$ smaller than the random stopping power. On the other hand, we note that the stopping
maximum for channeled ions is located at the same velocity as in the case of random ions, while 
in the case of a homogeneous electron gas with an electron density equal to the average electron
density along the channel would be located at a lower value of the velocity, as shown by the 
LDA calculation represented by a dotted line. We also note that within the LDA position-dependent
corrections to the random stopping power for slow ions are predicted to be too small; for velocities
above the stopping maximum, the local-density approximation yields unrealistic values for the energy
loss of best channeled ions, which are far below our full band-structure calculations. At high
velocities, position-dependent and random stopping powers are found to converge.

\section{Conclusions}

We have presented a survey of current investigations of the impact of band structure effects on
plasmon energies, plasmon linewidths, dynamic structure factors, and both random and
position-dependent stopping powers of Al and Si.

New {\it ab initio} calculations for the plasmon lifetime in Al and Si have been presented.
In the case of Al, we find the full width at half maximum to be in the limit as $q\to 0$
of $0.2{\rm eV}$, below the early calculation carried out by Sturm with use of the nearly free
electron approximation\cite{Sturm}. The plasmon linewidth
is found to increase as $q^2$ for small values of the momentum transfer ${\bf q}$, while it
increases very quickly near the plasmon cutoff, in agreement with the experiment.
A similar behaviour is found in the case of Si, though plasmon linewidths in this material
are found to be about five times wider than in Al.

As for the random stopping power, we find that it is, at low velocities, smaller in Si than
in Al, though they both have nearly the same valence electron density. At velocities just over
the plasmon threshold they are both below the stopping power of a FEG with $r_s\sim 2$. At
high velocities there are no band structure effects. The random stopping power of Si is found
to be, at low and intermediate velocities, about $10\%$
smaller than that of Al, in agreement with experimental measurements for either
protons\cite{Bauer} or antiprotons\cite{Moller}.

Differences between random and position-dependent stopping powers of Al are found to be up to
$10\%$ for projectiles incident in the (100) direction and up to  $20\%$ for projectiles
moving in the (111) direction. As for Si, the stopping power for best channeled ions
along the (110) direction  is found to be diminished with respect to the random stopping
power by about $20\%$ at velocities near the stopping maximum.

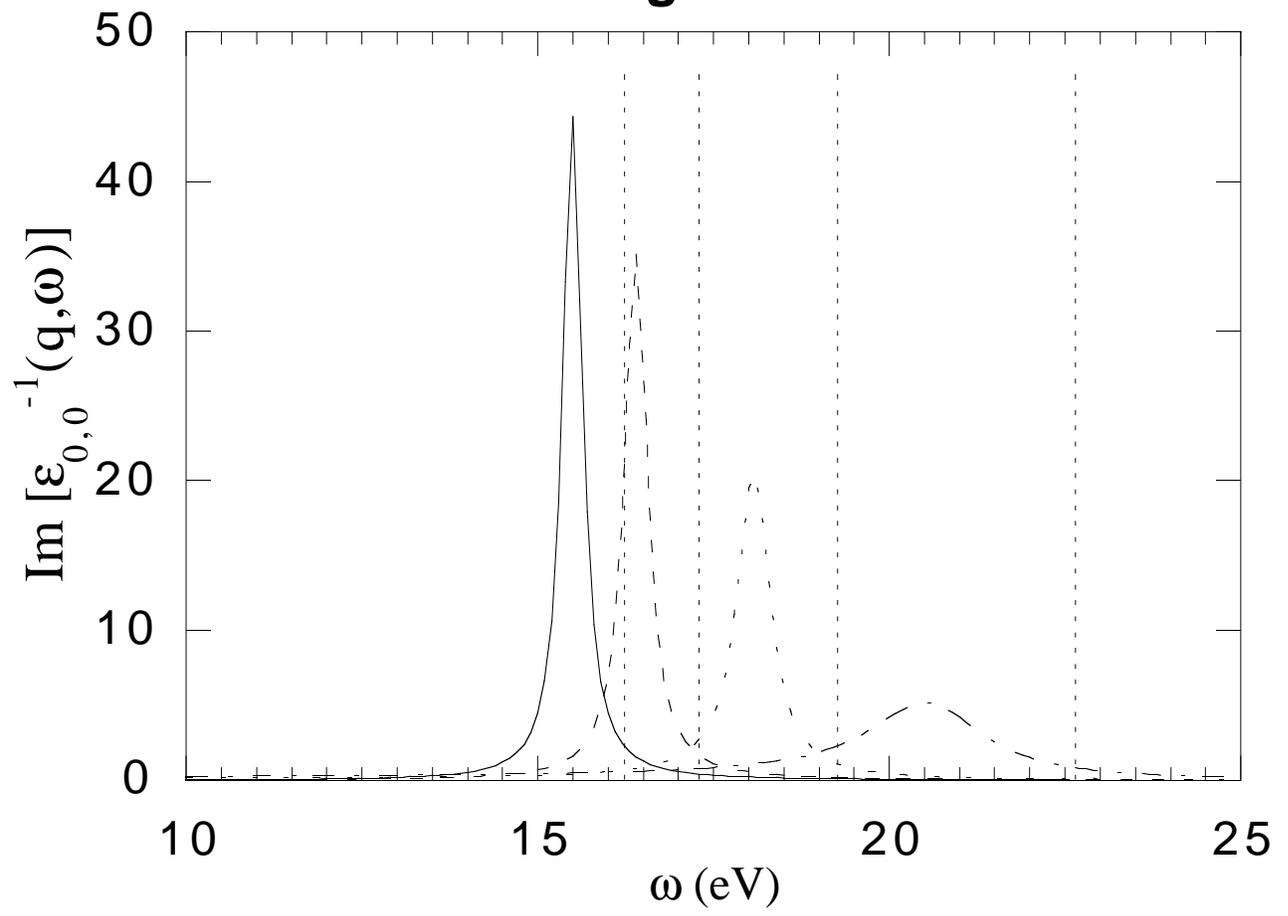
\begin{figure}
\caption{The RPA energy-loss function for Al, for several wave vectors along the (100) direction:
${\bf q}=(0.2,0,0){2\pi/a}$ (solid line), ${\bf q}=(0.4,0,0){2\pi/a}$ (long-dashed line),
${\bf q}=(0.6,0,0){2\pi/a}$ (short-dashed line), and ${\bf q}=(0.8,0,0){2\pi/a}$ (dashed-dotted
line). The vertical dotted lines represent the corresponding results for a FEG  with $r_s=2.07$,
which are simply $\delta$ functions.}
\end{figure}

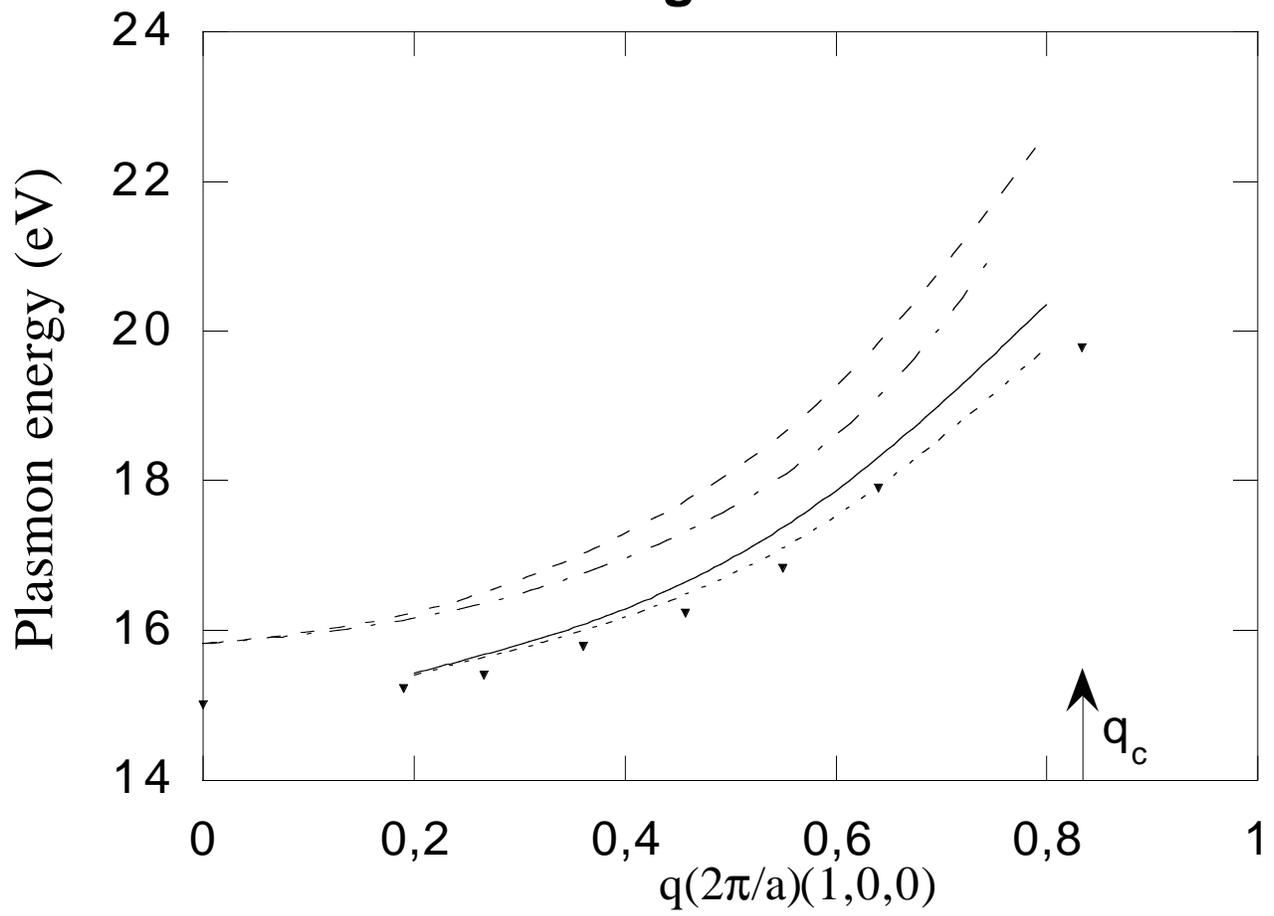
\begin{figure}
\caption{RPA (solid line) and TDLDA (dotted line) plasmon excitation energies for wave vectors
along the (100) direction in Al. The corresponding calculations for a FEG are represented
by dashed and dashed-dotted lines, as obtained within RPA and TDLDA, respectively. The experimental
results are represented by triangles}
\end{figure}

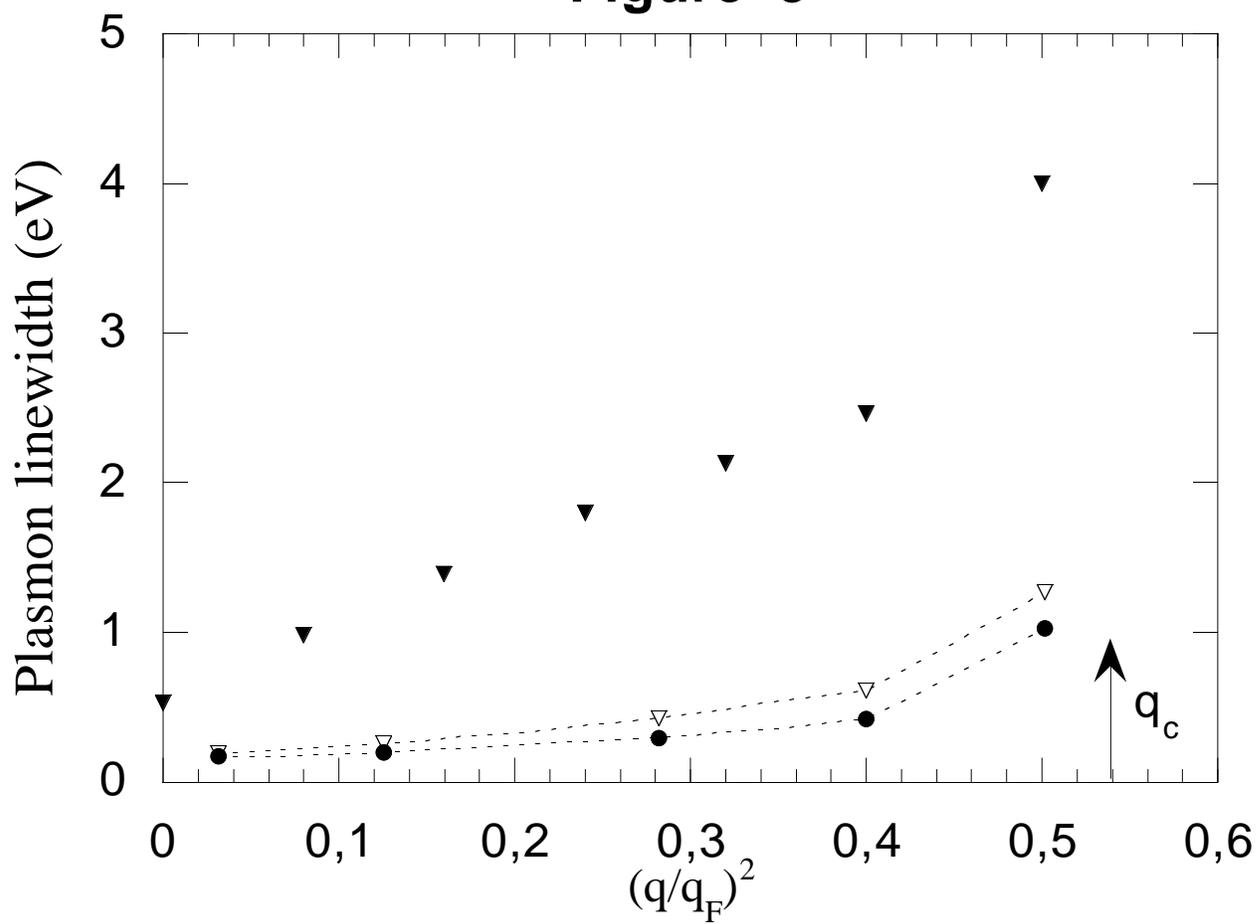
\begin{figure}
\caption{RPA plasmon linewidths of Al for wave vectors along the (100) direction. Solid circles
and open triangles represent calculated linewidths without and with inclusion of plasmon
decay through excitation of two electron-hole pairs. Solid triangles represent the experimental
results.}
\end{figure}

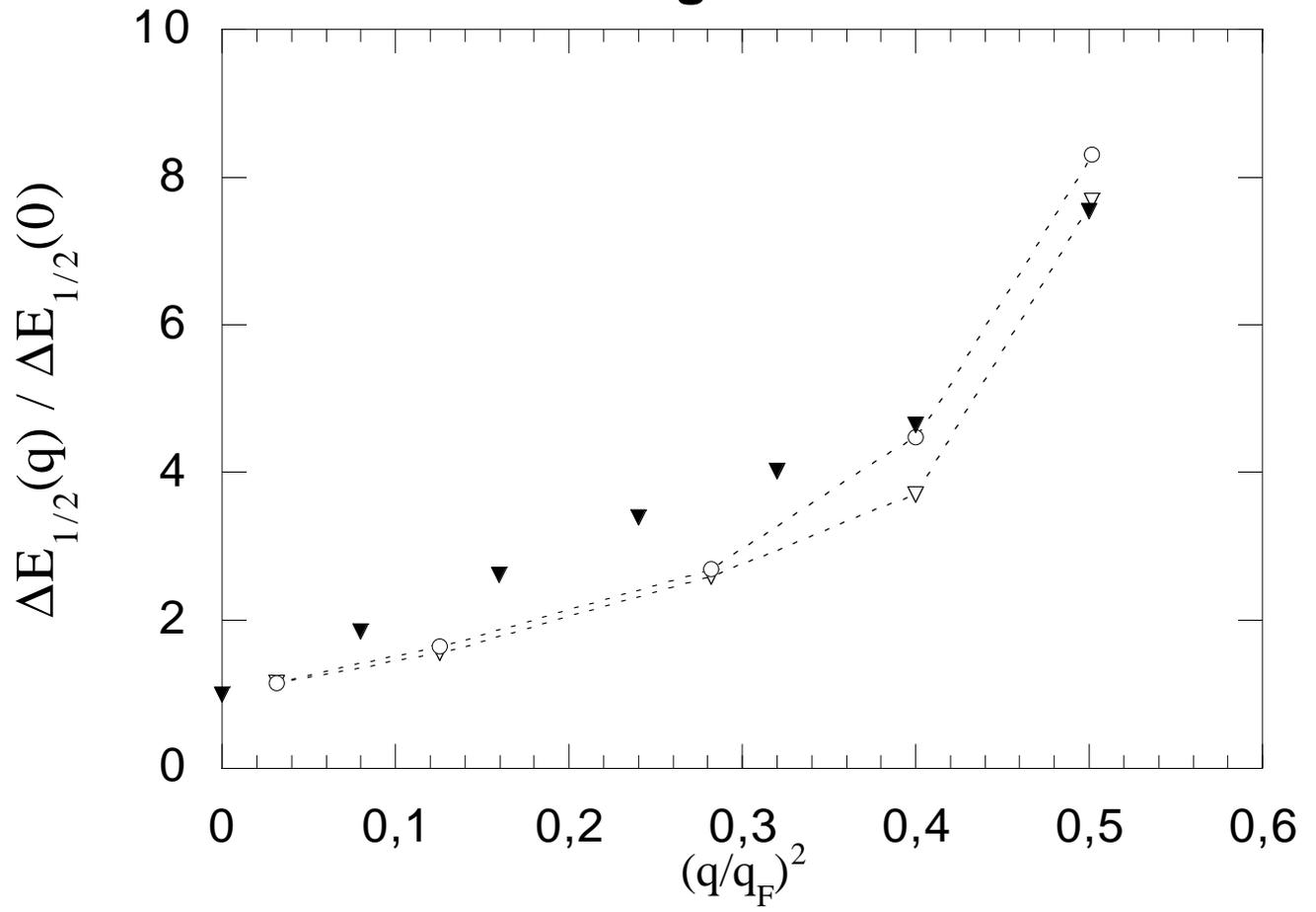
\begin{figure}
\caption{Scaled RPA (open triangles) and TDLDA (open circles) plasmon linewidths of Al for
wave vectors along the (100) direction, as obtained with inclusion of plasmon decay through
excitation of two electron-hole pairs. Solid triangles represent the experimental results.}
\end{figure}

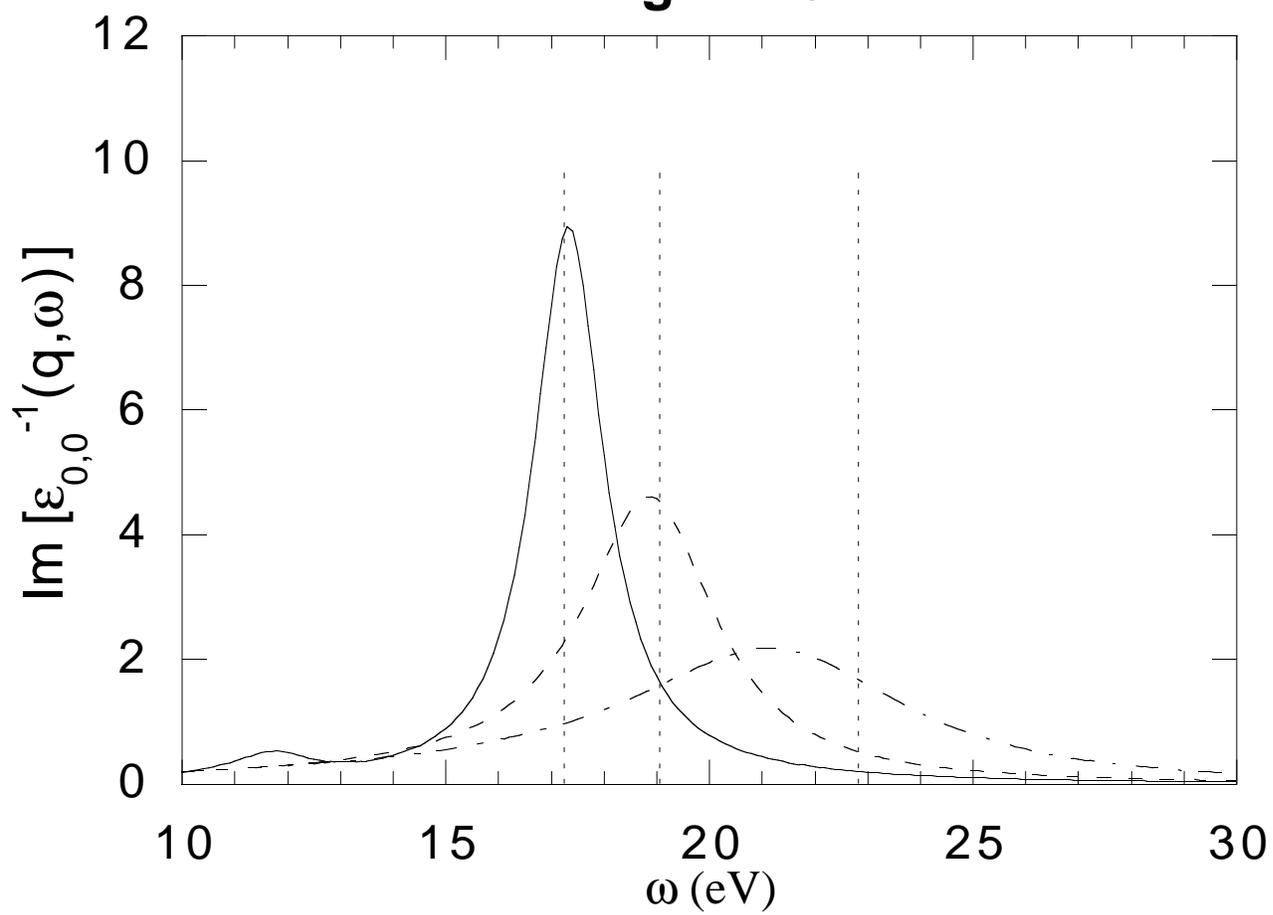
\begin{figure}
\caption{The RPA energy-loss function for Si, for several wave vectors along the (111) direction:
${\bf q}=(0.2,0.2,0.2){2\pi/a}$ (solid line), ${\bf q}=(0.4,0.4,0.4){2\pi/a}$ (dashed line), and
${\bf q}=(0.6,0.6,0.6){2\pi/a}$ (dashed-dotted line). The vertical lines represent the corresponding
results for a FEG with $r_s=2.01$, which are simply $\delta$ functions.}
\end{figure}

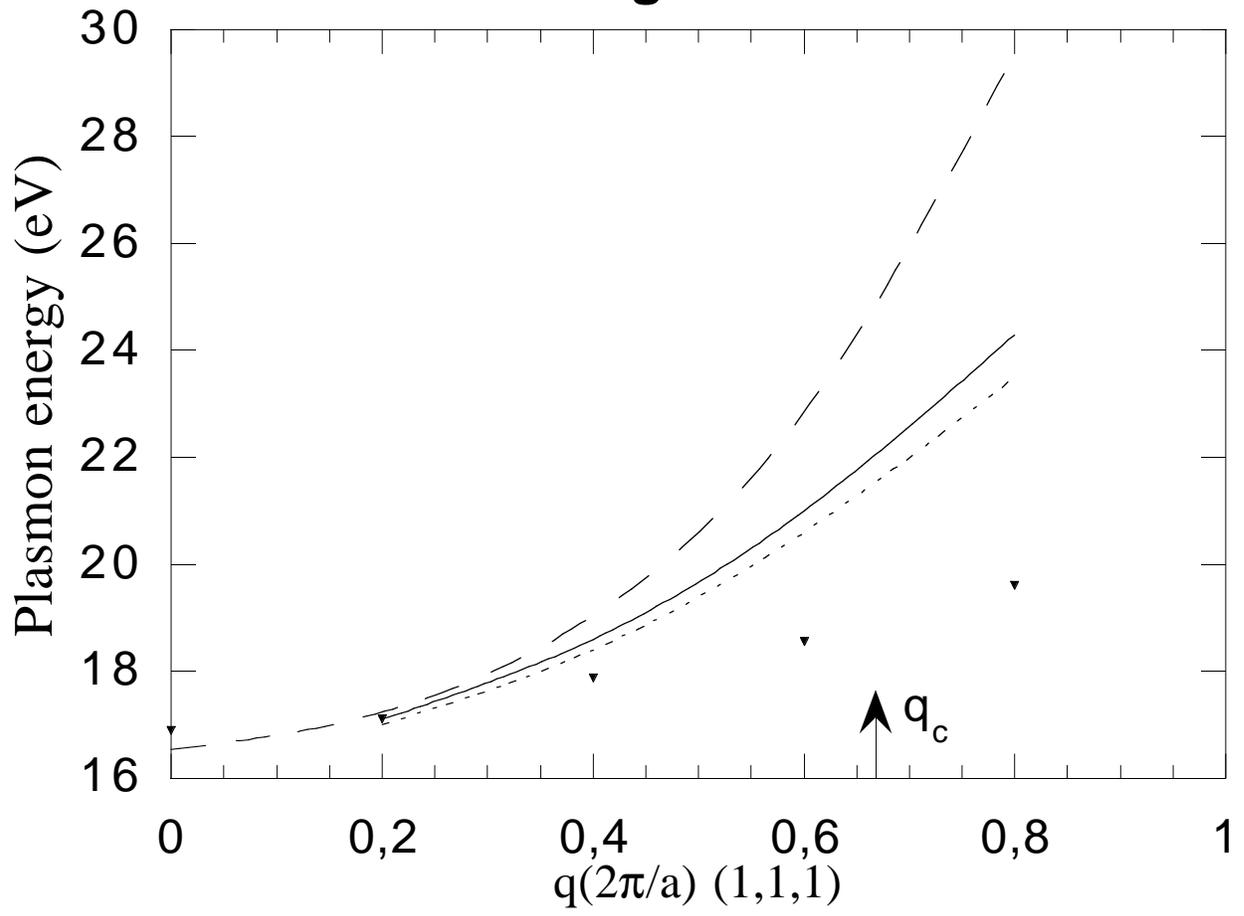
\begin{figure}
\caption{RPA (solid line) and TDLDA (dotted line) plasmon excitation energies for wave vectors
along the (111) direction in Si. The corresponding RPA calculations for a FEG are represented
by dashed lines. The experimental results are represented by triangles.}
\end{figure}

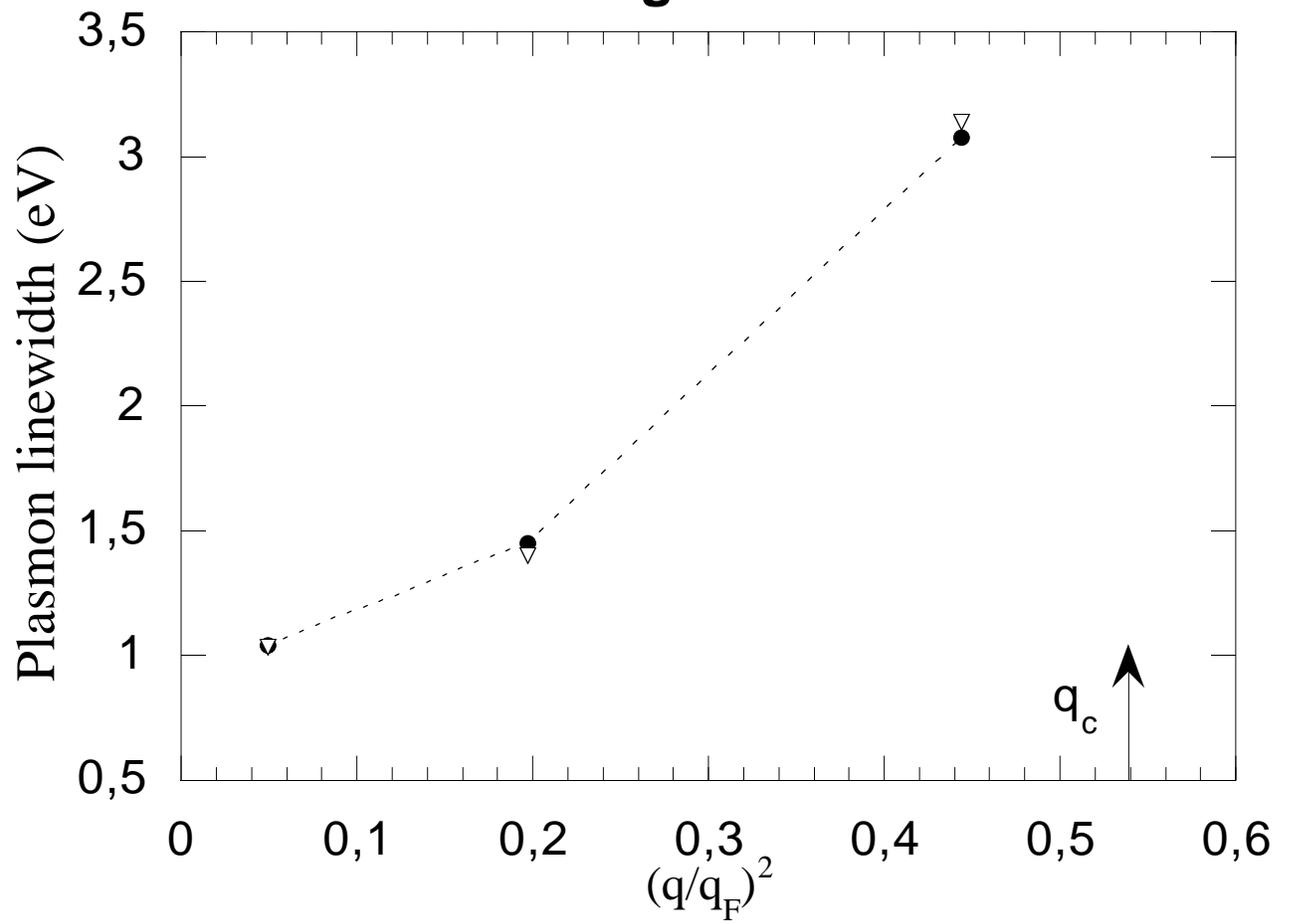
\begin{figure}
\caption{RPA (solid circles) and TDLDA (open triangles) plasmon linewidths of Si for wave vectors
along the (111) direction. Crystalline local-field effects have not been included in this
calculation.}
\end{figure}

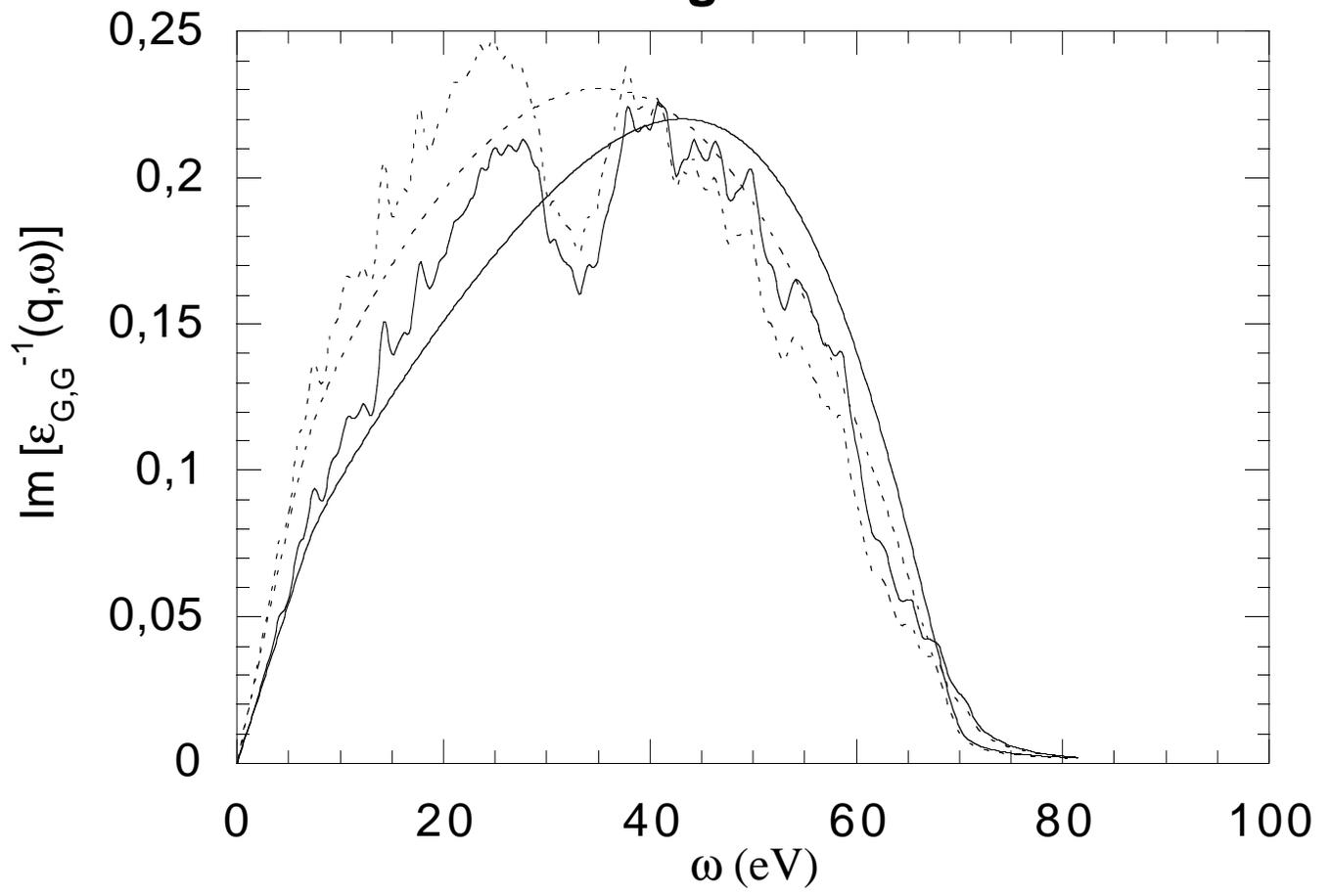
\begin{figure}
\caption{The RPA (solid lines) and TDLDA (dashed lines) energy-loss function for Al, for a large
wave vector of magnitude $|{\bf q}+{\bf G}|=1.65q_F$ along the direction (1,5/6,5/6). Plain solid
and dashed lines represent RPA and TDLDA energy-loss functions of a FEG with $r_s=2.07$.}
\end{figure}

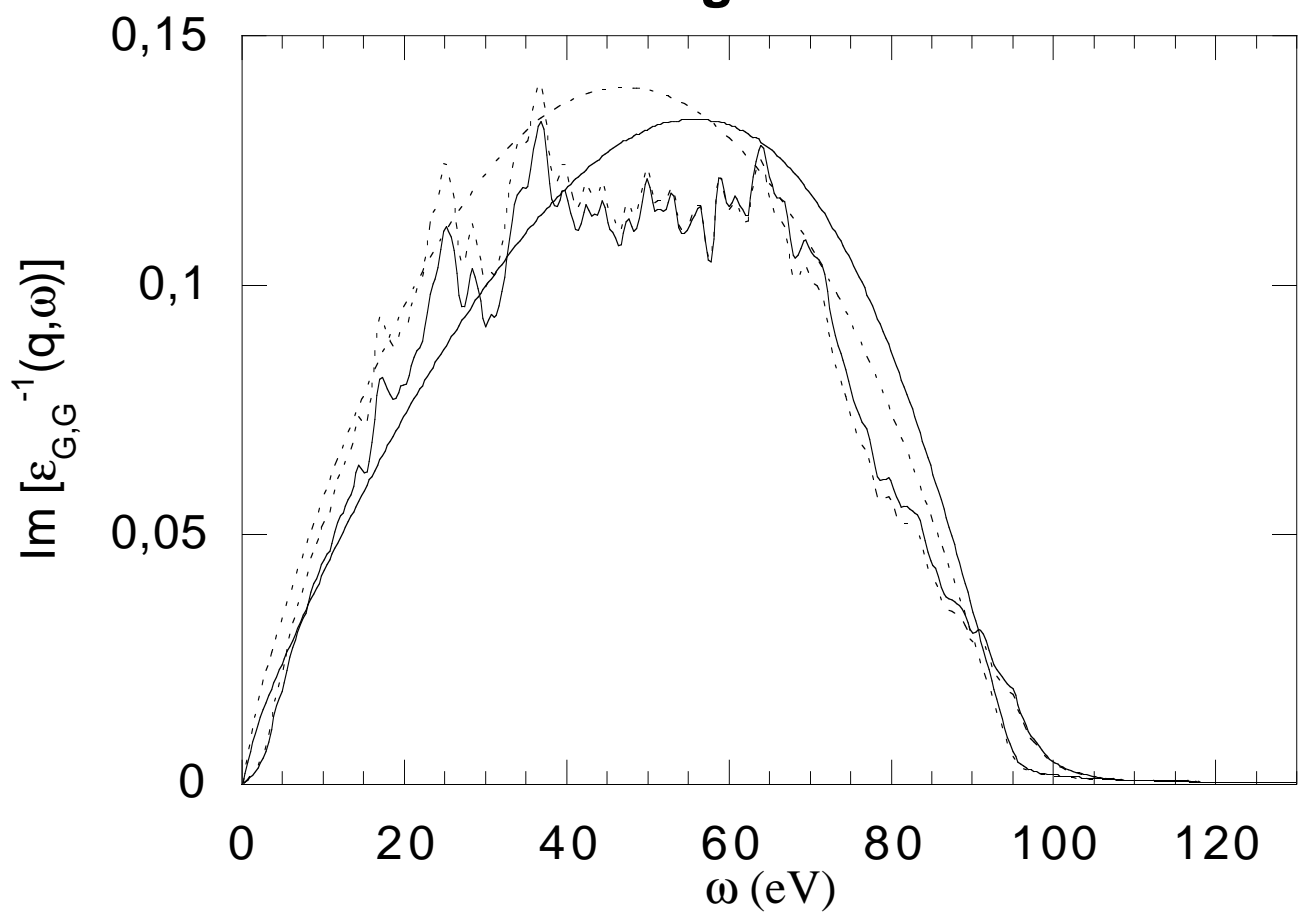
\begin{figure}
\caption{The RPA (solid lines) and TDLDA (dashed lines) energy-loss function for Si, for a large
wave vector of magnitude $|{\bf q}+{\bf G}|=1.95q_F$ along the direction (1,1,1). Plain solid
and dashed lines represent RPA and TDLDA energy-loss functions of a FEG with $r_s=2.01$.}
\end{figure}

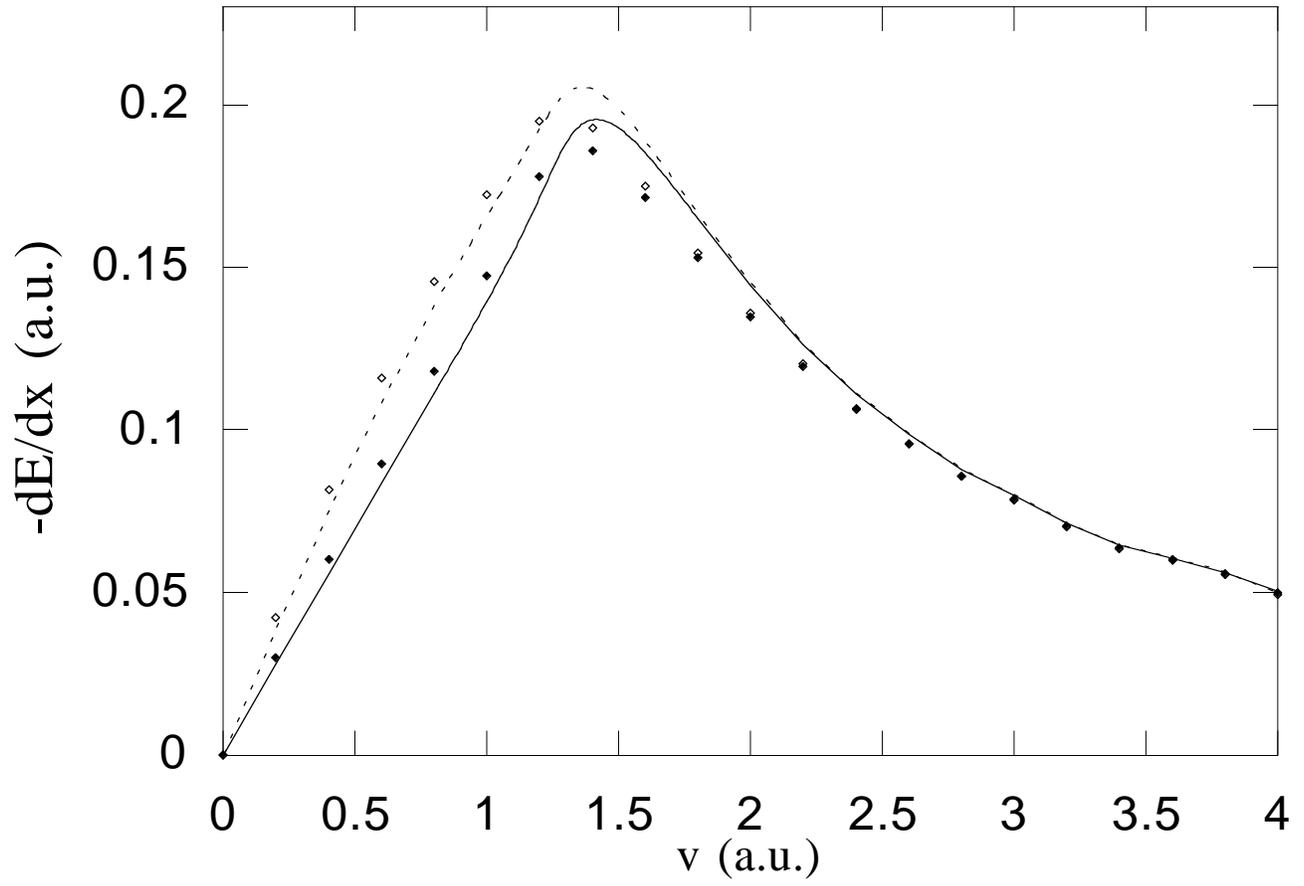
\begin{figure}
\caption{Full band-structure calculation of the random stopping power of valence electrons in Al
for protons ($Z_1=1$), as a function of the projectile velocity. Solid and open circles
represent the results obtained in the RPA and the TDLDA, respectively. RPA and TDLDA stopping powers
of electrons in a FEG with $r_s=2.07$ are represented by solid and dashed lines, respectively.}
\end{figure}

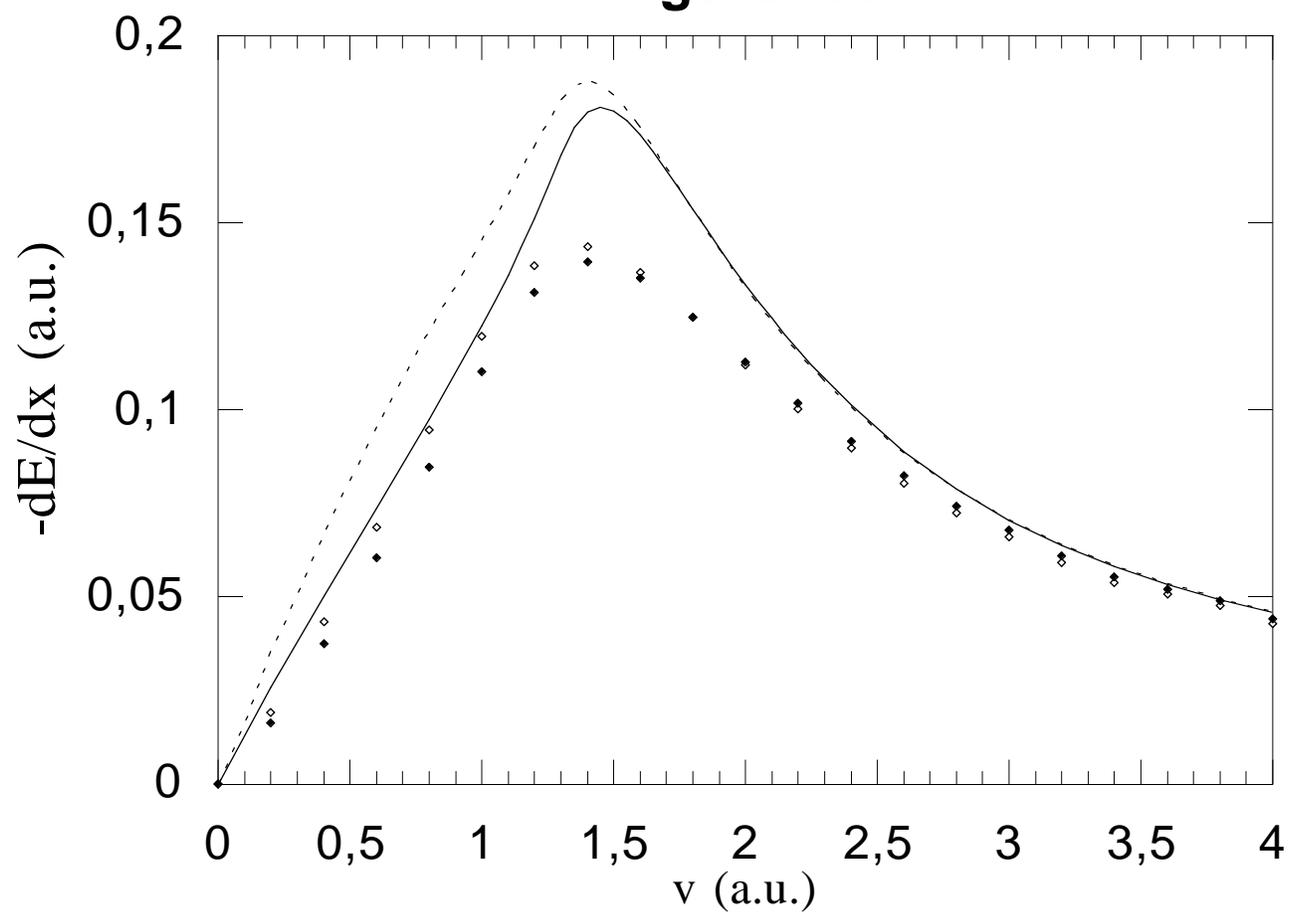
\begin{figure}
\caption{Full band-structure calculation of the random stopping power of valence electrons in Si
for protons ($Z_1=1$), as a function of the projectile velocity. Solid and open circles
represent the results obtained in the RPA and the TDLDA, respectively. RPA and TDLDA stopping powers
of electrons in a FEG with $r_s=2.01$ are represented by solid and dashed lines, respectively.}
\end{figure}

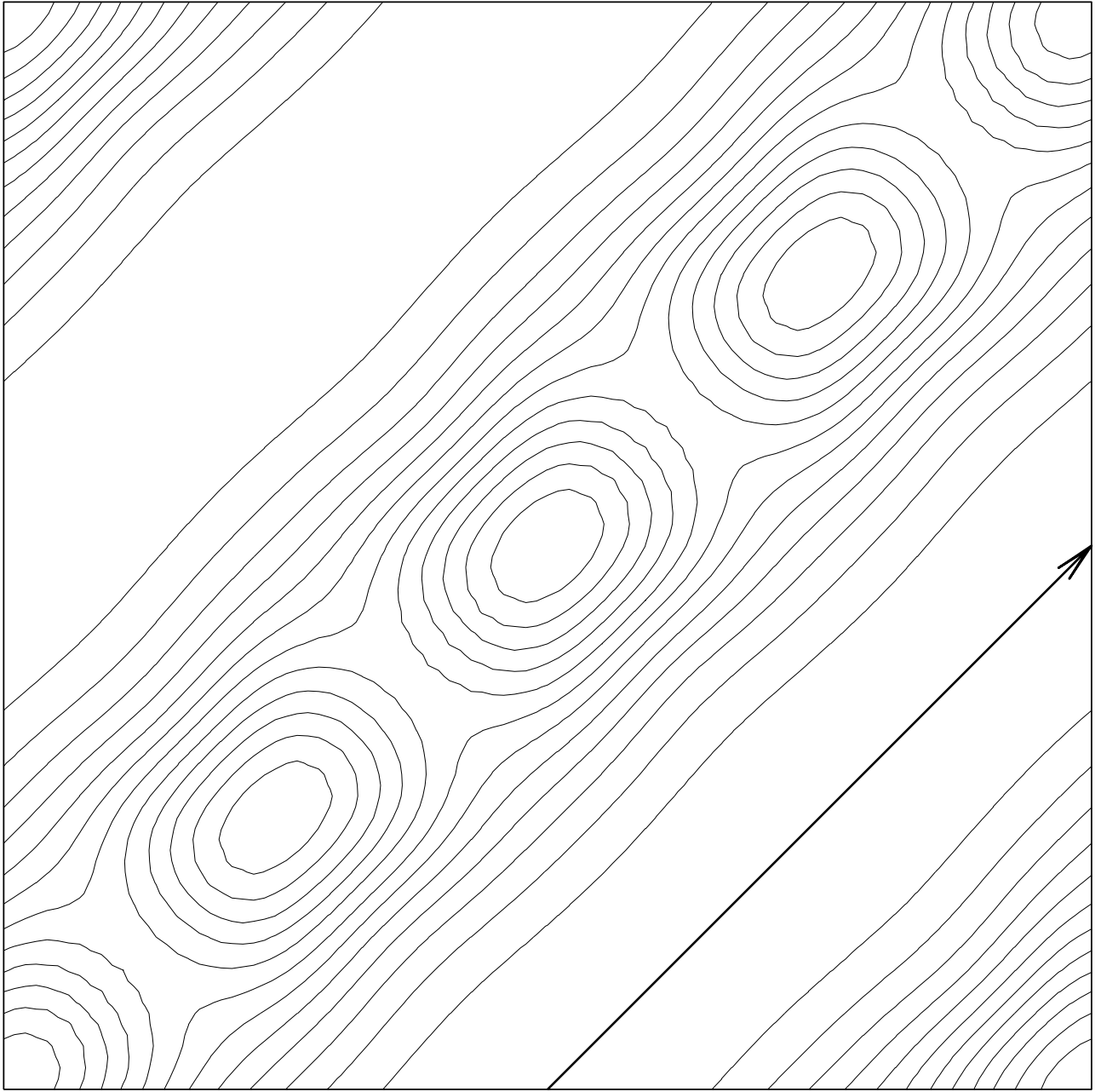
\begin{figure}
\caption{Contour density plot of the valence electron density in the plane defined by the
(100) and (010) vectors in Si.}
\end{figure}

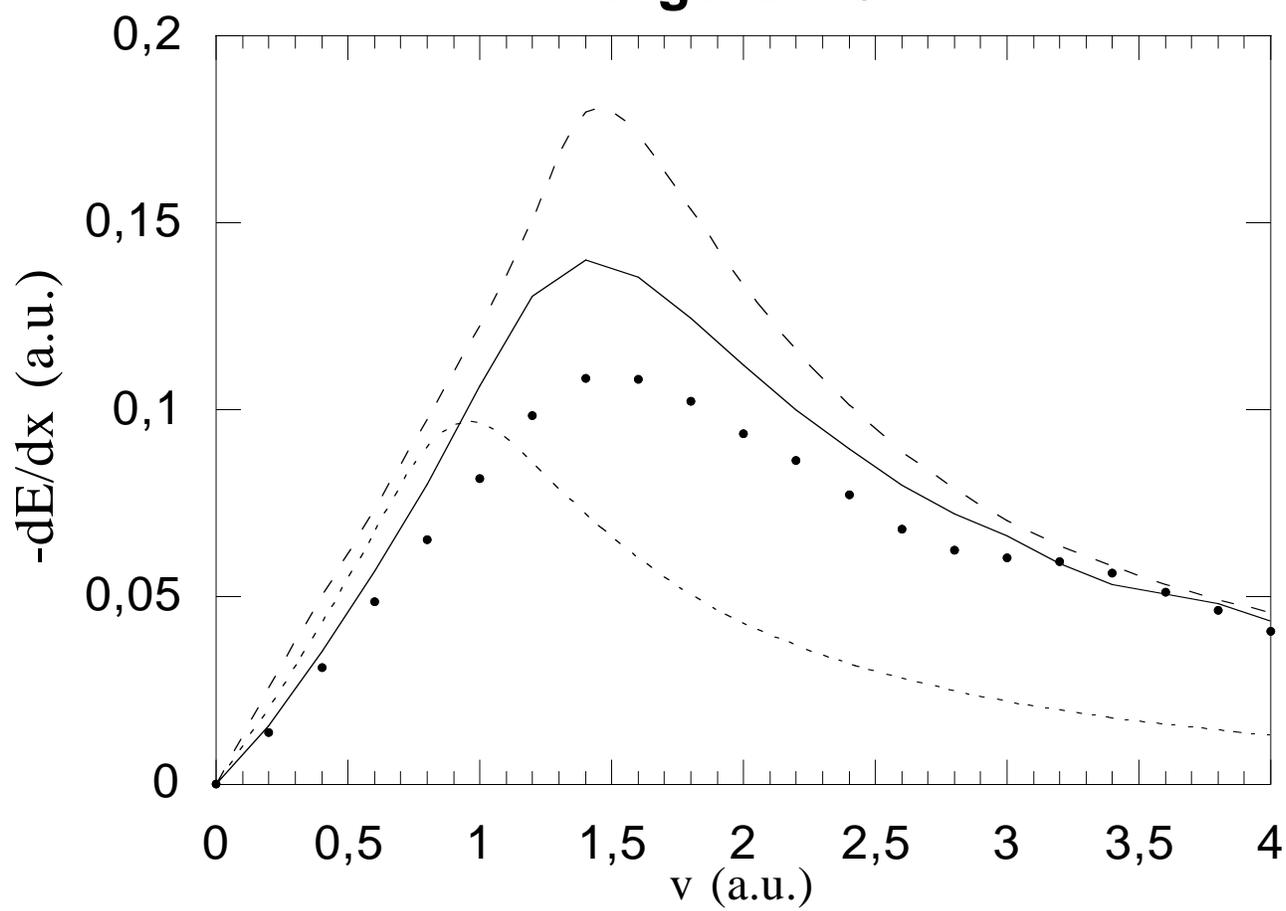
\begin{figure}
\caption{Full band-structure calculation of the RPA stopping power of valence electrons in Si
for best channeled protons ($Z_1=1$) in the (110) (solid circles), as a function of the 
projectile velocity. Solid and dashed lines represent the random stopping power of real Si
and of a FEG with $r_s=2.01$, respectively. The dotted line represents the stopping power
of a FEG with $r_s=3.37$.}
\end{figure}


\begin{references}
\bibitem{note} Both the stopping power of valence electrons in the real solid and that of a FEG have
been calculated with the integrations over the momentum transfer being extended from zero to $2.9q_f$
and $2.1q_F$ for Al and Si, respectively. Since non-zero contributions to the stopping power come,
within RPA, from all wave vectors that are smaller than $2(q_F+v)$, our calculated results are
expected to underestimate the total stopping power of both jellium and band electrons for nearly all
velocities under study. Nevertheless, we have found that differences, at large wave vectors, between
full and FEG calculations are small, and our results provide, therefore, an accurate measure of the
effect of the band structure on the random stopping power of valence electrons.
\end{references}
\end{document}